\documentclass{raa}            

\usepackage{graphicx,times}             
\usepackage{natbib}
\usepackage{amssymb,amsmath}
\bibpunct{(}{)}{;}{a}{}{,}

\usepackage[pagebackref=true]{hyperref}

\usepackage{tabularx}       
\usepackage{threeparttable}  
\usepackage{multicol}        
\usepackage{amsmath}         
\usepackage{adjustbox}
\usepackage{multirow}
\usepackage{array}
\usepackage{graphicx}
\usepackage[pagebackref=true]{hyperref}

\def\cm2{cm$^{-2}$}

\def\nh3{NH$_3$}
\def\n2h{N$_2$H$^+$}

\def\13co{$^{13}$CO}
\def\c18o{C$^{18}$O}
\def\hc3n{HC$_3$N}
\def\h2{H$_2$}
\def\nh{n(H$_2$)}

\begin{document}

  \title{The FAST Discovery of a binary millisecond pulsar PSR~J1647-0156B (M12B) with a candidate cross matching algorithm
}

   \volnopage{Vol.0 (202x) No.0, 000--000}      
   \setcounter{page}{1}          

    \author{Qiuyu Yu 
      \inst{1,2}
   \and Yujie Wang
      \inst{3,4}
   \and Zhichen Pan
      \inst{5,6,7,8}
   \and Zhongli Zhang
      \inst{3,4}
   \and Lei Qian
      \inst{5,6,7,8}
   \and Zhongzu Wu
      \inst{1}
     \and Ralph~P.~Eatough
     \inst{5}
   \and Dejiang Yin 
      \inst{1} 
   \and Baoda Li 
      \inst{1}   
   \and Yujie Chen
      \inst{1} 
   \and Yinfeng Dai 
      \inst{9,10}   
   \and Yifeng Li
      \inst{11,12} 
   }

 \institute{College of Physics, Guizhou University, Guiyang 550025, China;{\it zzwu08@gmail.com}\\
       \and
           Guizhou Vocational College of Foodstuff Engineering, Guiyang 550025, China;\\
        \and
            Shanghai Astronomy Observatory, Chinese Academy of Sciences, Shanghai 200030, China;{\it zzl@shao.ac.cn}\\
        \and
            School of Astronomy and Space Sciences, University of Chinese Academy of Sciences, Beijing 100049, China;\\
        \and
            National Astronomical Observatories, Chinese Academy of Sciences, Beijing 100012, P.R. China; {\it panzc@nao.cas.cn;lqian@nao.cas.cn}\\
        \and
            Guizhou Radio Astronomical Observatory, Guizhou University, Guiyang 550025, P.R. China;\\
        \and
             College of Astronomy and Space Sciences, University of Chinese Academy of Sciences, Beijing 100101, P.R. China;\\
        \and
             Key Laboratory of Radio Astronomy, Chinese Academy of Sciences, Beijing 100101, P.R. China;\\
        \and     
             School of Physics and Astronomy, Beijing Normal University, Beijing 100875, China;\\
        \and
             Department of Physics, Faculty of Arts and Sciences, Beijing Normal University, Zhuhai 519087, China;\\
        \and     
             National Time Service Center, Chinese Academy of Sciences, Xi‘an 100101, China;\\
        \and
             University of Chinese Academy of Sciences, Beijing 100049, China\\
   }

\abstract{ We propose a pulsar candidate cross matching algorithm to sift radio pulsar search candidates from repeated observations of the same sky location such as globular clusters, high energy sources, or supernova remnants. 
Our method uses both the candidate spin period($P$) and dispersion measure(DM) value; 
if two or more candidates from different observations have similar spin periods to within 1\%, and dispersion measure values within 10\%, they are likely to correspond to the same candidate detection.
We have demonstrated the effectiveness of our method through the discovery of the pulsar M12B with the Five-hundred-meter Aperture Spherical radio Telescope (FAST). This pulsar has a spin period of 2.76\,ms and a dispersion measure of $42.70 \pm 0.05\,\mathrm{cm}^{-3}~\mathrm{pc}$. This pulsar has a profile with three peaks, being faint, showing scintillation. It is in an approximately 0.53-day orbit. Our discovery indicates that more pulsars might be effectively discovered if the algorithm is applied to the search results from other archival 
globular cluster observations.
}

   \titlerunning{The FAST Discovery of a binary millisecond pulsar PSR~J1647-0156B (M12B) }  

   \maketitle

%
%
\begin{multicols}{2}
\section{Introduction}           
\label{sect:intro}

The common pulsar search methods are typically based on the Fast Fourier Transform (FFT)\citep{2002AJ....124.1788R}, the single-pulse search method(\citealt{2003ApJ...596.1142C};\citealt{2003ApJ...596..982M}), for the detection of isolated dispersed pulses, and/or the Fast Folding Algorithm (FFA)\citep{2020MNRAS.497.4654M}.

The FFT exploits pulsar periodicity by converting time-domain data to frequency-domain power spectra.
The corresponding pulsar search identifies peaks at the pulsar fundamental/harmonic frequencies. Single pulse searches work by convolving boxcar filters of different widths throughout the dedispersed time series, and searching for peaks.
The FFA directly folds dedispersed time-series at trial periods into phase-averaged pulse profiles,
with detection determined by maximizing the Signal-to-Noise Ratio (SNR) of the pulse profile.

While all of these methods have been essential in the discovery of new pulsars and transient sources, they all face a critical challenge, i.e., the dramatic expansion in the number of pulsar candidates from the search analysis of large-scale surveys with highly sensitive new radio telescopes such as FAST\citep{2011IJMPD..20..989N}.
The FAST Globular Cluster (GC) pulsar survey has produced a large volume of archival data and search results(\citealt{2025ApJS..279...51L}). Many GCs hosting known pulsars have been observed multiple, often dozens, of times. Standard pulsar search procedures analyze each observation separately and apply fixed SNR thresholds to select pulsar candidates for visual inspection. Typically, weak pulsar signals, or those with unusual pulse profiles, may be missed during visual inspection. Although machine learning solutions to the problem are becoming increasingly common \citep{2014MNRAS.443.1651M}, logistical challenges to training and validation sets for new pulsar surveys remain \citep{2024ChJPh..90..121C}.

In this work, we developed a Cross Matching Algorithm (hereafter termed CMA) that filters candidates from multiple observations of the same source. By identifying candidates with similar $P$ and DM, but detected at different observational epochs, we can now extract candidates based on their multiplicity, helping detecting faint or unusual candidates that can be missed in independent analyses. Additionally the algorithm reduces the volume of search candidates to be visually inspected, easing the workload on observers and mitigating the chance to miss pulsars.


\section{Methods}
\label{methods}

Firstly, the CMA combines the sifted candidate lists of different epochs, generated by {\sc PRESTO}\citep{2001PhDT.......123R,2002AJ....124.1788R}) {\tt ACCEL\_sift.py} into a single list. 

Secondly, Each candidate from any observational epoch serves as a reference and is compared with all candidates from all epochs to form a subgroup. A candidate pair is considered a match only if both of the following criteria are satisfied: (1)the  relative dispersion measure (DM) difference $\Delta \mathrm{DM}/\mathrm{DM} \le 10\%$, where the DM value is from the reference candidate.  (2) the relative spin period difference $\Delta P/P\le 1\%$, where $P$ is from the reference candidate. These matches are then grouped with a hash mapping dictionary, where each unique candidate pair is assigned a distinct hash key\citep{2025ITIF...20.6635L}. In practice this was implemented in our CMA script\footnote{\url{https://github.com/1994yuqiuyu/cross-matching-for-pulsars}} by using {\tt defaultdict} in the {\tt collections} module of {\sc python}

We evaluated the impact of the criteria on candidate sifting efficiency. 
Adopting more strict thresholds results in a reduced number of candidates per group but carries the risk of missing pulsars. 
Conversely, looser criteria lead to larger candidate lists with the contamination by RFI. 
After comparative testing of various tolerance combinations, 
we found that a 1\%  difference for spin period and 10\% difference for DM represents a balanced configuration, facilitating the maximum recovery of potential pulsar candidates.

\begin{figure*}[htbp]
   \centering
   \includegraphics[width=1\textwidth]{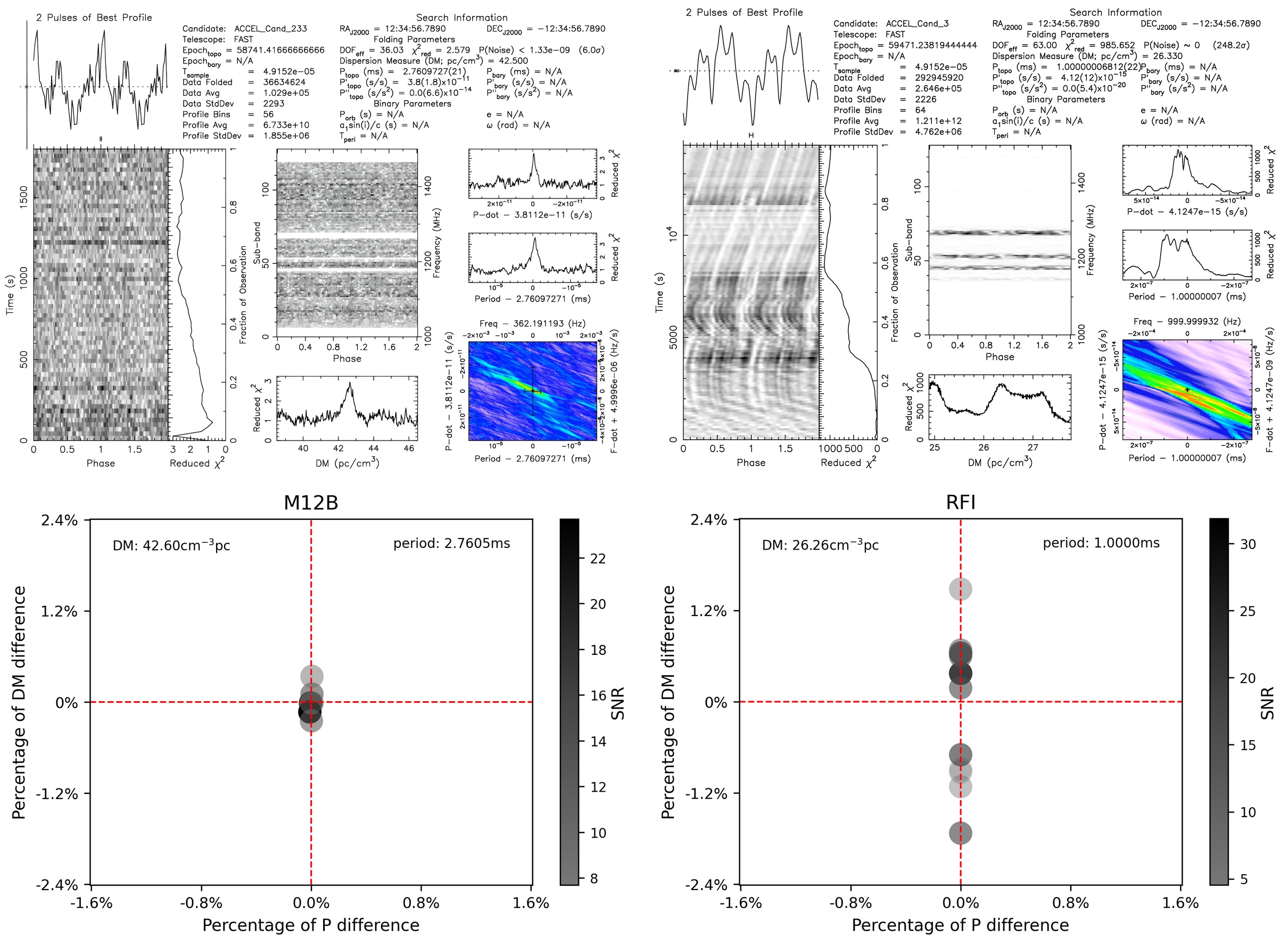}
   \caption{Diagnostic plot. 
Left upper: The detection plots of M12B. 
Lower left: Diagnostic scatter plot of M12B candidates in the P–DM parameter space. Each point’s horizontal coordinate represents the percentage deviation of the candidate’s measured period $P_i$ from the arithmetic mean period $\bar{P}$ of the M12B candidate group, calculated as: $\frac{P_i - \bar{P}}{\bar{P}} \times 100\%$. The vertical coordinate denotes the percentage deviation of the candidate’s measured dispersion measure $\mathrm{DM}_i$ from the group’s arithmetic mean $\overline{\mathrm{DM}}$, computed as: $\frac{\mathrm{DM}_i - \overline{\mathrm{DM}}}{\overline{\mathrm{DM}}} \times 100\%$. The origin (0\%, 0\%) in this panel corresponds to the group-average reference point $(\bar{P}, \overline{\mathrm{DM}})$, with all M12B candidate points plotted relative to this mean value. 
Right upper:the detection plots of RFI.
Right lower:the diagnostic plot of RFI, displaying the distribution of RFI candidates in the identical P–DM percentage deviation parameter space as M12B for direct comparative analysis.}
   \label{fig:Fig1}
   \end{figure*}

Third, a diagnostic plot of $\Delta \mathrm{DM}/\mathrm{DM}$ vs. $\Delta P/P$, is created to distinguish CMA pulsar signals from radio frequency interference (RFI) or candidates from random superpositions (Figure~\ref{fig:Fig1}).
In the left half of Figure~\ref{fig:Fig1}, an individual epoch detection of M12B (the pulsar discovery in this work - see Section~\ref{dataprocdisc}) 
is shown alongside a typical repeated RFI signal
(the right half of Figure~\ref{fig:Fig1}). The upper panels show standard {\sc presto} {\tt prepfold} candidate diagnostic plots. The lower panels show the new diagnostic plot for the CMA showing detections clustering in the parameter space of DM, $P$ and SNR. The genuine pulsar discovery (M12B) displays less distribution in detected $P$ and DM, whereas RFI shows significant scatters in DM. There are more examples of both genuine pulsar and RFI detections found with the CMA.\footnote{\url{https://cstr.cn/31253.11.sciencedb.j00167.00029}}.

As the last step, all matched candidate diagnostic plots were visually inspected to ensure the consistent pulse profile and DM from different observations.

As a test, we checked the CMA on detections of known pulsars in a variety of observations 
from the FAST GC pulsar survey\citep{2025ApJS..279...51L}. The involved GCs are M2, M3, M12, M15, and NGC~6517
The thresholds for DM and $P$ were varied while keeping all the known pulsar signals detectable,

The known pulsar detection details, including the number of detection, standard deviation of $P$, standard deviation of DM, the relative deviation  $\left\vert\frac{\overline{P}-P_0}{P_0}\right\vert$ and $\left\vert\frac{\overline{\text{DM}}-\text{DM}_0}{\rm DM_0}\right\vert$ can be found in Table \ref{tab:sifted_result}, where $\overline{\text{DM}}$ and $\overline{P}$ are the average of the DM and $P$, respectively. DM$_0$ and $P_0$ are the published dispersion measure and spin period, respectively.

\begin{table*}[!ht]
    \centering
    \small
    \caption{Summary of pulsar search results in globular clusters. $N$: number of candidate detections from the same pulsar sifted by the CMA; $\sigma_P$: standard deviation of $P$ (scaled by $\times1000$); $\sigma_{\mathrm{DM}}$: standard deviation of DM; $\overline{P}$ and $\overline{\mathrm{DM}}$: average period and DM of the CMA-associated candidate group; $P_0$ and $\mathrm{DM}_0$: published reference values. References:1.\citep{2021ApJ...915L..28P}, 2.\citep{2025ApJS..279...51L}, 3.\citep{2024ApJ...972...43L}, 4.\citep{2024ApJ...974L..23W}, 5.\citep{2025RAA....25g1001D}, 6.\citep{2024ApJ...969L...7Y}}
    \label{tab:sifted_result}
    \begin{tabularx}{\linewidth}{@{}lccccccccc@{}}
    \hline
   Pulsar Name & $N$ & $\sigma_P$ & $\sigma_{\mathrm{DM}}$ & $\overline{P}$ & $\overline{\mathrm{DM}}$ & $P_0$ & $\mathrm{DM}_0$ & $\frac{\overline{P}-P_0}{P_0}$ & $\frac{\overline{\mathrm{DM}}-\mathrm{DM}_0}{\mathrm{DM}_0}$ \\
    & &(ms) & (pc cm$^{-3}$) & (ms) & (pc cm$^{-3}$) & (ms) & (pc cm$^{-3}$) & (\textperthousand) & (\%) \\
    \hline
    M2A & 7 & 0.8 & 0.1 & 10.1 & 43.4 & 10.1 & 43.4 & 0.1 & 0.3 \\
    M2B & 8 & 1.1 & 0.1 & 7.0 & 43.8 & 7.0 & 43.7 & 0.7 & 0.1 \\
    M2C & 14 & 2.0 & 0.2 & 3.0 & 44.1 & 3.0 & 44.0 & 1.7 & 0.1 \\
    M2D & 6 & 0.2 & 0.1 & 4.2 & 43.5 & 4.2 & 43.6 & 1.0 & 0.2 \\
    M2E & 5 & 3.8 & 0.2 & 3.8 & 43.6 & 3.7 & 43.7 & 1.6 & 0.5 \\
    M2F & 2 & 0.5 & 0.1 & 4.8 & 43.3 & 4.8 & 43.3 & 0.2 & 0.3 \\
    M3B & 4 & 0.2 & 0.1 & 2.4 & 26.1 & 2.4 & 26.1 & 0.0 & 0.3\\
    M3D & 12 & 0.2 & 0.1 & 5.4 & 26.3 & 5.4 & 26.4 & 0.0 & 0.1 \\
    M3F & 2 & 0.0 & 0.0 & 4.4 & 26.4 & 4.4 & 26.4 & 0.1 & 0.1 \\
    M12A & 26 & 2.8 & 0.2 & 2.4 & 42.4 & 2.4 & 42.5 & 1.2 & 0.2 \\
    M12B & 9 & 0.2 & 0.1 & 2.8 & 42.7 & 2.8 & 42.7 & 0.3 & 0.1 \\
    M15A & 13 & 4.0 & 0.2 & 110.7 & 67.1 & 110.7 & 67.2 & 0.0 & 0.1 \\
    M15B & 12 & 1.8 & 0.1 & 56.1 & 67.8 & 56.1 & 67.7 & 0.0 & 0.1 \\
    M15D & 27 & 4.3 & 0.2 & 4.8 & 67.2 & 4.8 & 67.3 & 0.8 & 0.1 \\
    M15E & 12 & 1.1 & 0.1 & 4.7 & 66.6 & 4.7 & 66.6 & 0.1 & 0.0 \\
    M15F & 15 & 5.4 & 0.3 & 4.0 & 65.8 & 4.0 & 65.6 & 0.4 & 0.2 \\
    M15G & 3 & 1.3 & 0.4 & 37.7 & 67.0 & 37.7 & 66.4 & 0.0 & 1.0 \\
    M15H & 10 & 3.7 & 0.3 & 6.7 & 67.2 & 67.7 & 67.1 & 0.3 & 0.1 \\
    M15I & 10 & 7.1 & 0.5 & 5.1 & 67.5 & 5.1 & 67.5 & 1.6 & 0.0 \\
    M15J & 9 & 0.4 & 0.1 & 11.8 & 66.7 & 11.8 & 66.7 & 0.0 & 0.0 \\
    M15M & 10 & 6.3 & 0.4 & 4.8 & 67.9 & 4.8 & 67.9 & 0.6 & 0.1 \\
    NGC6517A & 10 & 0.6 & 0.3 & 7.2 & 182.6 & 7.2 & 182.7 & 0.1 & 0.0 \\
    NGC6517B & 9 & 1.3 & 0.0 & 29.0 & 182.4 & 29.0 & 182.4 & 0.1 & 0.0 \\
    NGC6517C & 15 & 4.8 & 1.8 & 3.7 & 183.5 & 3.7 & 182.4 & 1.0 & 0.6 \\
    NGC6517D & 10 & 0.9 & 0.3 & 4.2 & 174.5 & 4.2 & 174.5 & 0.1 & 0.0 \\
    NGC6517E & 9 & 0.4 & 0.1 & 7.6 & 183.3 & 7.6 & 183.2 & 0.0 & 0.1 \\
    NGC6517F & 9 & 1.2 & 0.0 & 24.9 & 183.9 & 24.9 & 183.8 & 0.0 & 0.1 \\
    NGC6517G & 8 & 2.4 & 0.1 & 51.6 & 185.2 & 51.6 & 185.1 & 0.0 & 0.1 \\
    NGC6517H & 10 & 2.4 & 0.3 & 5.6 & 179.7 & 5.6 & 179.6 & 0.1 & 0.1 \\
    NGC6517I & 12 & 2.9 & 0.5 & 3.3 & 177.6 & 3.3 & 177.9 & 0.8 & 0.2 \\
    NGC6517K & 6 & 2.5 & 0.2 & 9.6 & 182.4 & 9.6 & 182.4 & 0.2 & 0.0 \\
    NGC6517L & 7 & 2.1 & 0.3 & 6.1 & 185.9 & 6.1 & 185.7 & 0.0 & 0.1 \\
    NGC6517M & 7 & 0.1 & 0.2 & 5.4 & 183.3 & 5.4 & 183.2 & 0.0 & 0.1 \\
    NGC6517N & 11 & 2.9 & 0.8 & 5.0 & 182.9 & 5.0 & 182.6 & 0.6 & 0.1 \\
    NGC6517O & 6 & 2.3 & 0.7 & 4.3 & 182 & 4.3 & 182.5 & 0.2 & 0.3 \\
    NGC6517P & 16 & 4.4 & 0.7 & 5.5 & 183.8 & 5.5 & 183.5 & 0.7 & 0.4 \\
    \hline
    \end{tabularx}
    \footnotesize
\end{table*}

\section{Data Processing and the New Pulsar Discovery}
\label{dataprocdisc}
We used a collection of 64 FAST observations of the GCs M2, M3, M12, M15, and NGC~6517. 
The observation settings and data reduction details can be found in the related papers 
\citep{2021ApJ...915L..28P, 2021RAA....21..143P, 2024ApJ...969L...7Y, 2024ApJ...972...43L, 2024ApJ...974L..23W}.
At the end of the search pipeline, the standard pulsar search software (e.g., \textsc{PRESTO}) produces a sifted list of candidates
for all DM and acceleration trials, each containing a large number (approximately 500 to 1500) of candidates that are each `folded' and inspected visually.

The coordinates of these GCs, the observation dates, durations, the number of candidates from every observational data set, the number of candidates after CMA, and the known pulsar detections can be found in Table \ref{tab:combined_observations}. Six (M2A to F) of 10 known pulsars were identified in M2.
The pulsar M2G to I only appeared in the candidate lists once in all the observations, while M2J was not detected in all the observations.

\begin{table*}[thp]
\centering
\caption{Observation data details and processing results. The observation duration in seconds $(T_{\rm obs})$ is given in brackets underneath the observation date; $N_{\rm cand}$ specifies the number of candidates produced by {\tt ACCEL\_sift.py}; \textbf{T:} Total of $N_{{\rm cand}}$ ; \textbf{P:} The number of candidates that were detected at least twice by the CMA ($\Delta P/P \leq 1\%$, $\Delta\mathrm{DM}/\mathrm{DM} \leq 10\%$); \textbf{K:} Known pulsar detections (confirmed/total in cluster); \textbf{N:} New pulsars discovered. The main reason for missing some known pulsars is that these pulsars were listed as candidates in the observations for no more than once }
\label{tab:combined_observations}
\begin{adjustbox}{width=\textwidth, center}
\scriptsize
\begin{tabular}{p{2cm}p{0.9cm}p{1.2cm}p{1.6cm}cp{1.6cm}cp{1.6cm}c p{1.3cm}}
\hline
GC Name & \begin{tabular}{@{}c@{}}R.A.\\J2000\end{tabular} & \begin{tabular}{@{}c@{}}Decl.\\J2000\end{tabular} & \begin{tabular}{@{}c@{}}Date\\$(T_{\rm obs}\,(\rm{s}))$\end{tabular} & \begin{tabular}{@{}c@{}}$N_{\rm cand}$\end{tabular} & \begin{tabular}{@{}c@{}}Date\\$(T_{\rm obs})$\end{tabular} & \begin{tabular}{@{}c@{}}$N_{\rm cand}$\end{tabular} & \begin{tabular}{@{}c@{}}Date\\$(T_{\rm obs})$\end{tabular} & \begin{tabular}{@{}c@{}}$N_{\rm cand}$\end{tabular} & \begin{tabular}{@{}c@{}}Results\end{tabular} \\
\hline
\multirow{8}{*}{M2 (NGC~7089)} & \multirow{8}{*}{21:33:27} & \multirow{8}{*}{-00:49:23.5} 
& 2021-09-29 (7188) & 744 & 2021-10-03 (7200) & 733 & 2021-10-11 (7200) & 825 & \multirow{8}{*}{\begin{tabular}{@{}c@{}}T: 8996\\P: 349\\K: 6/10\\N: 0\end{tabular}} \\
& & & 2021-10-27 (7158) & 669 & 2021-12-07 (7200) & 700 & 2021-12-20 (7200) & 679 & \\
& & & 2022-01-16 (7188) & 755 & 2022-01-25 (7200) & 1292 & 2022-03-24 (7395) & 1268 & \\
& & & 2022-07-04 (6798) & 703 & 2022-07-28 (7985) & 628 & & & \\
\hline

\multirow{16}{*}{M3 (NGC~5272)} & \multirow{16}{*}{13:42:12} & \multirow{16}{*}{+28:22:38.2} 
& 2020-08-28 (1800) & 51 & 2020-09-23 (7200) & 98 & 2020-09-29 (7200) & 63 & \multirow{16}{*}{\begin{tabular}{@{}c@{}}T: 2151\\P: 331\\K: 3/5\\N: 0\end{tabular}} \\
& & & 2020-10-08 (7200) & 76 & 2020-10-27 (14400) & 115 & 2020-11-30 (18000) & 88 & \\
& & & 2020-12-05 (5880) & 82 & 2020-12-11 (18000) & 92 & 2020-12-21 (11115) & 79 & \\
& & & 2021-01-12 (18015) & 118 & 2021-01-17 (2715) & 87 & 2021-01-25 (18015) & 96 & \\
& & & 2021-01-26 (18015) & 74 & 2021-02-05 (14160) & 103 & 2021-02-14 (14400) & 79 & \\
& & & 2021-02-15 (18015) & 87 & 2021-03-18 (18015) & 94 & 2021-03-19 (18015) & 114 & \\
& & & 2021-03-20 (18015) & 109 & 2021-03-21 (19184) & 80 & 2021-07-14 (16377) & 101 & \\
& & & 2021-12-08 (14400) & 101 & 2022-04-18 (5040) & 112 & 2022-06-04 (4380) & 52 & \\
\hline

\multirow{6}{*}{M12 (NGC~6218)} & \multirow{6}{*}{16:47:14} & \multirow{6}{*}{-01:56:54.7} 
& 2019-10-30 (10200) & 6067 & 2021-04-18 (6380) & 1876 & 2021-09-26 (12871) & 1669 & \multirow{6}{*}{\begin{tabular}{@{}c@{}}T: 10599\\P: 413\\K: 2/2\\N: 1\end{tabular}} \\
& & & 2022-06-08 (4680) & 318 & 2022-06-09 (5400) & 604 & 2022-07-02 (3720) & 320 & \\
& & & 2022-12-03 (7200) & 240 & & & & & \\
\hline

\multirow{10}{*}{M15 (NGC~7078)} & \multirow{10}{*}{21:29:58} & \multirow{10}{*}{+12:10:01.2} 
& 2019-11-09 (4800) & 2351 & 2020-08-30 (3300) & 2925 & 2020-09-04 (1800) & 4516 & \multirow{10}{*}{\begin{tabular}{@{}c@{}}T: 36009\\P: 4874\\K: 10/15\\N: 0\end{tabular}} \\
& & & 2020-09-22 (9000) & 2434 & 2020-09-25 (14400) & 1285 & 2020-10-15 (300) & 6340 & \\
& & & 2022-11-19 (6600) & 1849 & 2023-02-20 (6600) & 1688 & 2023-12-19 (15840) & 2369 & \\
& & & 2024-01-22 (9060) & 2179 & 2024-01-23 (10800) & 2231 & 2024-01-30 (6320) & 2801 & \\
& & & 2024-02-17 (6375) & 3041 & & & & & \\
\hline

\multirow{6}{*}{NGC~6517} & \multirow{6}{*}{18:01:51} & \multirow{6}{*}{-08:57:31.6} 
& 2020-01-19 (7200) & 1185 & 2020-01-20 (7200) & 1073 & 2020-01-21 (7200) & 1063 & \multirow{6}{*}{\begin{tabular}{@{}c@{}}T: 12442\\P: 1549\\K: 15/21\\N: 0\end{tabular}} \\
& & & 2020-01-22 (4800) & 1436 & 2020-01-23 (9000) & 1616 & 2020-09-13 (7200) & 1485 & \\
& & & 2020-04-30 (7200) & 529 & 2022-12-28 (8400) & 1971 & 2022-12-31 (8400) & 2084 & \\
\hline
\end{tabular}
\end{adjustbox}
\end{table*}

The pulsar M15M was very faint and detected by stacking spectra from different observations (Dai et al. in prep).
With the CMA, it was detected, too.
It appeared in candidate lists for 10 times and was ignored previously because it is too faint.

\begin{figure*}[htbp]
  \centering
      \includegraphics[width=0.497\textwidth]{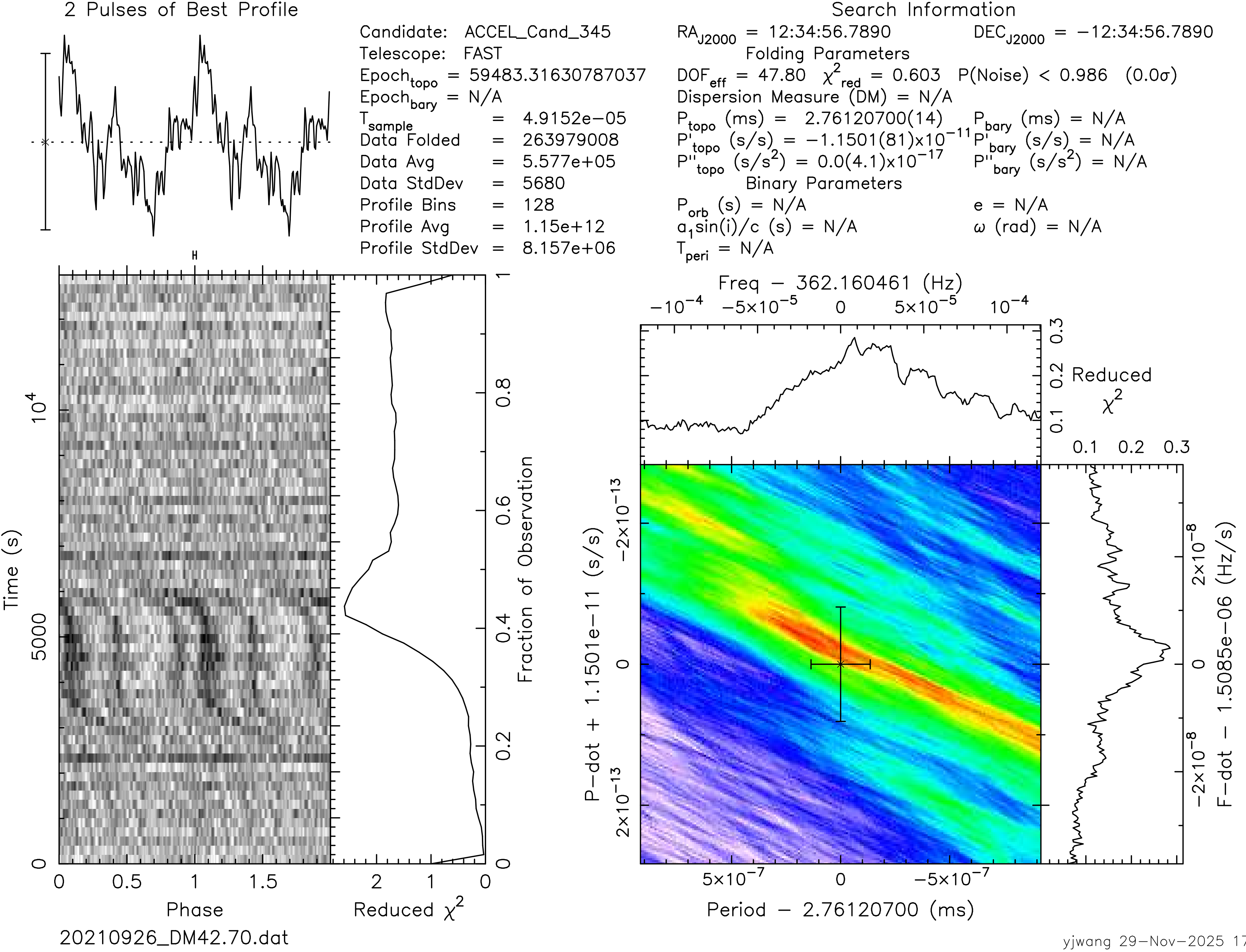}
  \includegraphics[width=0.497\textwidth]{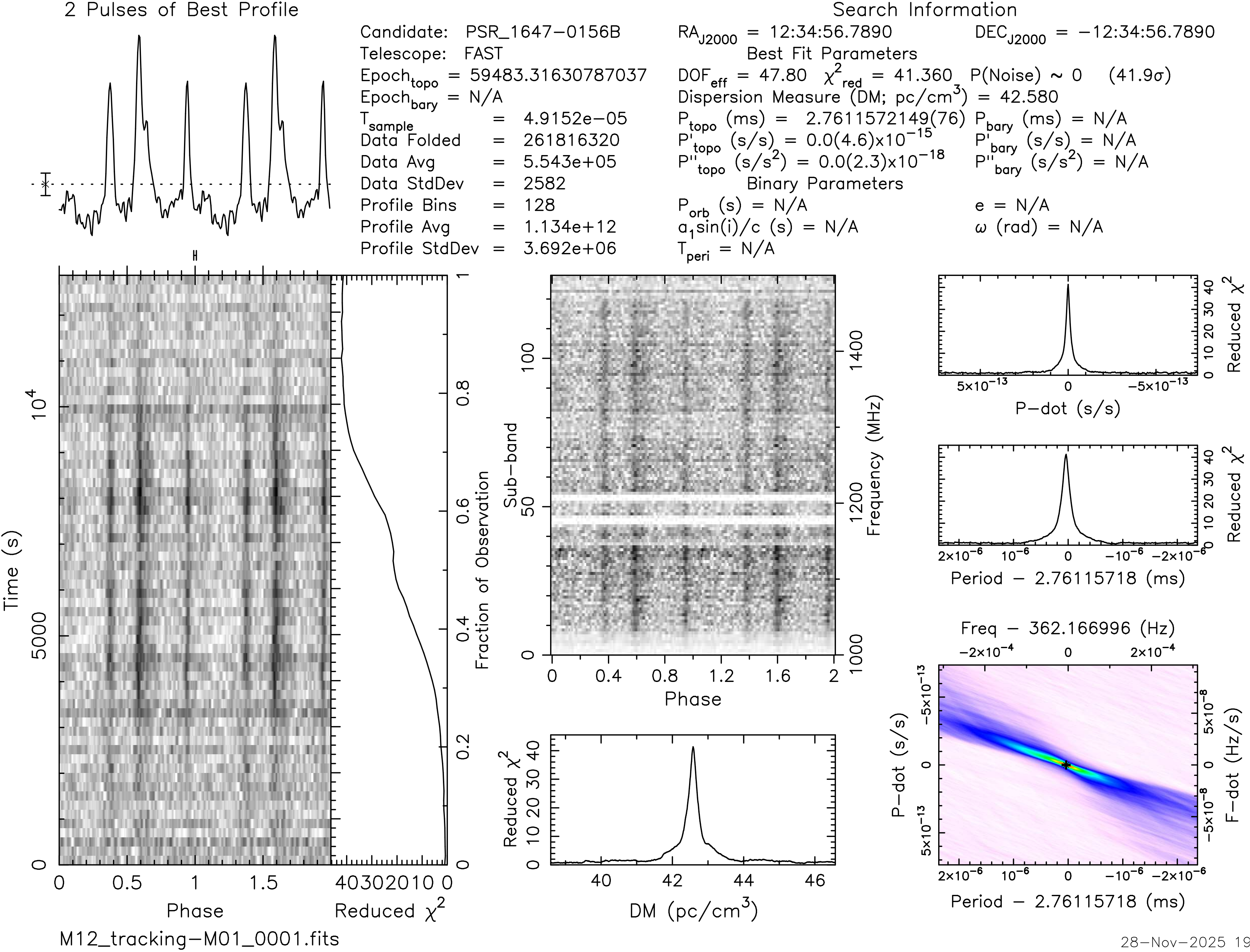} 
  \caption{
Left:the detection plots of M12B. Right: the M12B signal from folding with partially timing solution, showing three peaks in the pulse profile.
  }
  \label{fig:M12B}
\end{figure*}

In M12, only the pulsar M12A was known previously. In this work, a new pulsar, namely M12B (PSR~J1647-0156B) was identified.
It was listed as a candidate for 9 times.
Its DM values ranged from 42.55 to 42.80 $\mathrm{cm}^{-3}~\mathrm{pc}$, while its spin periods ranged between 2.7605 and 2.7609 ms.
The previously discovered binary millisecond pulsar (MSP) M12A (PSR~J1648-0156A) has a DM value of 42.50 $\mathrm{cm}^{-3}~\mathrm{pc}$, which is close to that of M12B, indicating that both pulsars are likely the members of M12.
The discovery plots of M12B are presented in Figure~\ref{fig:M12B}. 
When folding the candidate in different observations, the pulse profile is consistent with each other. 
M12B has three relatively narrow peaks in the pulse profile. It also shows scintillation.
Initial timing solutions suggest that the system has a nearly circular orbit ($e$ $\approx$ 0) with an orbital period $P_\mathrm{b} \approx$ 0.53 days and a projected semi-major axis $x$ $\approx$ 0.73 lt-s. 
The estimated minimum and median companion masses are 0.15 and 0.18 $M_{\odot}$, respectively, assuming a pulsar mass of 1.4 $M_{\odot}$. 
These parameters fall within the range expected for low-mass binary MSPs formed from low-mass X-ray binaries (LMXBs)\citep{2011ASPC..447..285T}.

Once the M12B was discovered, the reason for its non-detection in previous observations became clear.
In a 0.5-day orbit, it was easily missed in the searches with the full dataset, which lasted for at least 2 hours. 

It can be detected in segmented searches, which divided the data into 0.5- and 1-hour segments.
However, in these detections, the pulsar was too faint to be considered as a real one. 
With the CMA, these vague detections were collected, 
allowing the pulsar to be sifted and further identified.

\section{Discussion and Conclusion}
\label{discussion}

The thresholds of either spin period or DM mainly came from the tests with known pulsars.
The 10\% difference in DM values comes mainly from experiences in candidate selection and confirmation.
If the difference is set between 5\% and 15\%, the results are almost the same.
If the difference in DM values is 18\% or larger, RFIs would be mixed into the pulsar signals.
The intrinsic DM of a pulsar can be considered as a constant.
The observed DM values may vary due to fluctuations in the interstellar medium and the noises.
Luckily, from our results, for a detected signal, 
the spin periods (1\% difference) and DM values (10\% difference) are not significantly affected.

\begin{figure*}[htbp]
  \centering
  \includegraphics[width=0.65\textwidth]{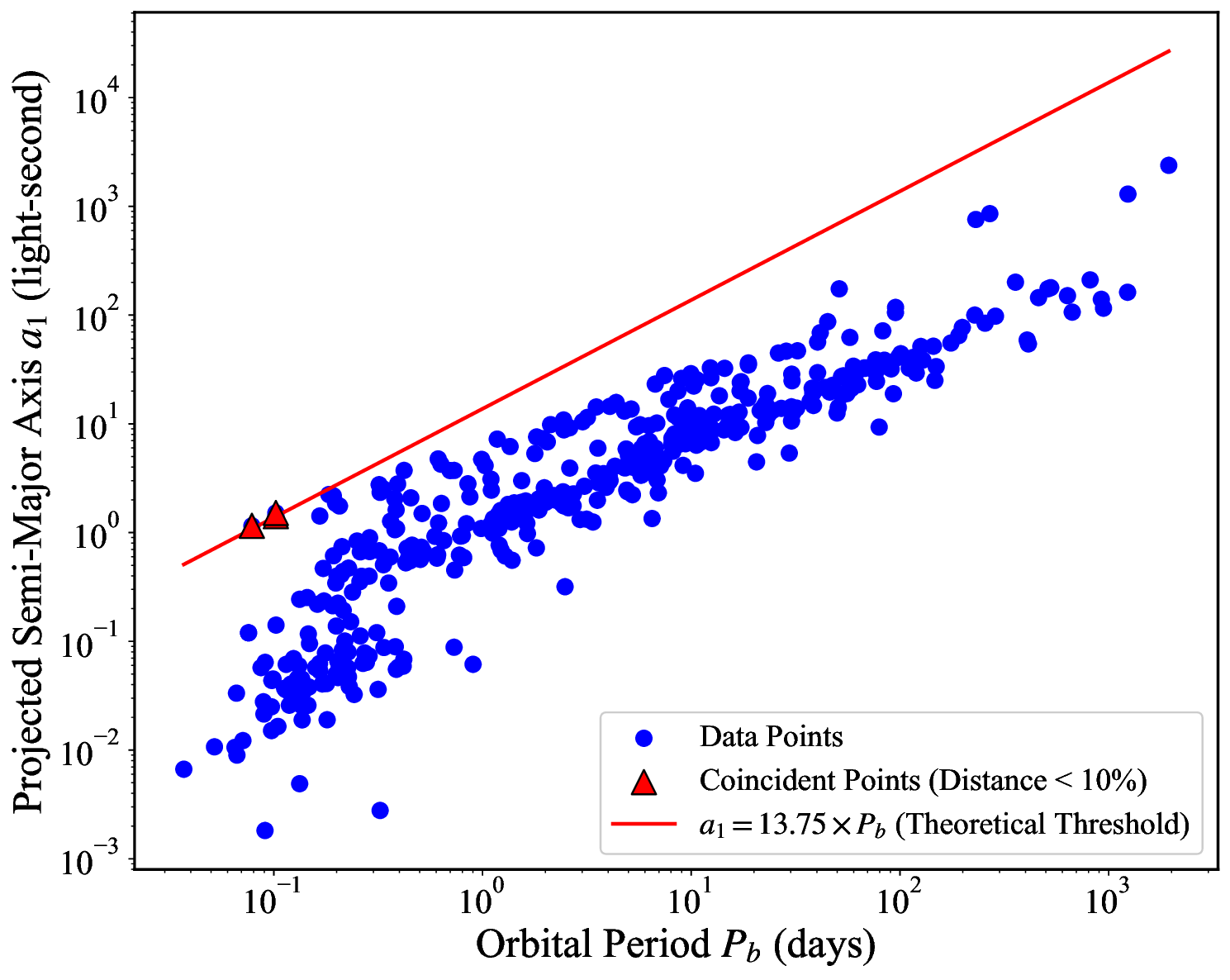}
  \caption{The distribution of orbital periods and semi-major axes of known binary pulsars.
  The double neutron star PSR~J0737-3039A and B, and PSR~J1946+2052 were near the red line which is
  for the equation $a_1 = \dfrac{0.01 \cdot c \cdot P_{\mathrm{b}}}{2\pi}$ that both of them may have 1\% variation in spin periods due to the orbital movement.
  For other pulsars, the 1\% upper limit is enough for the $\Delta P/P$.
  }
  \label{fig:a1_pb}
\end{figure*}

The main factor causing the period variation (the threshold of $\Delta P/P \leq 1\%$) is the orbital motion of a binary pulsar (see Appendix A). The orbital periods and semi-major axes of 438 binary pulsars are obtained from {\sc psrcat}\footnote{https://www.atnf.csiro.au/research/pulsar/psrcat/}. Figure~\ref{fig:a1_pb}  displays the distribution of orbital periods and semi - major axes for known binary pulsars.
Two binary systems (PSR J0737-3039A/B and PSR J1946+2052) are close to the limits.
Both of them are double neutron star systems (massive) and locate in very short orbit.
Consequently, the condition $\Delta P/P_0 \leq 1\%$ holds universally for all known pulsars. 
Besides, additional RFI showed up when applying period difference ratios of 1.5\% and 2\% in the CMA.

Currently, we cannot obtain the final phase-connected timing solution for M12B due to the insufficient data. 
A tentative phase connected timing solution from  4 observations and with a 10-day duration shows an extremely large period derivative (around $10^{-10}$ s/s level),  indicating a  multiple-companion nature of M12B.

Based on the studies in this work, the conclusions are the follows:

1. CMA can sift faint and/or scintillating candidates from multiple observations with similar spin periods and DM values, with the suggested thresholds of $\Delta P/P \le 1\%$ and $\Delta \mathrm{DM}/\mathrm{DM} \le 10\%$, respectively.

2. A new pulsar M12B (PSR~J1647-0156B) was discovered.
It has a spin period of 2.76 ms and a DM of $42.70 \pm 0.05\,\mathrm{cm}^{-3}~\mathrm{pc}$.
The DM of M12B is quite close to that of M12A ( $42.50 \pm 0.05\,\mathrm{cm}^{-3}~\mathrm{pc}$), 
indicating that both are the members of M12.
M12B locates in a binary system with an orbital period of $\rm \sim$0.5 days.
Currently, we cannot obtain the phase-connected timing solution for M12B due to insufficient data.

\begin{acknowledgements}

This work is supported by the National Key R\&D Program of China No. 2022YFC2205202, No. 2020SKA0120100 and the National Natural Science Foundation of China (NSFC, Grant Nos. 12373032, 12003047, 11773041, U2031119, 12173052, and 12173053.
Both Lei Qian and Zhichen Pan were supported by the Youth Innovation Promotion Association of CAS (id.~2018075, Y2022027 and 2023064), and the CAS "Light of West China" Program.
This work made use of the data from FAST (Five-hundred-meter Aperture Spherical radio Telescope) (https://cstr.cn/31116.02.FAST). 
FAST is a Chinese national mega-science facility, operated by National Astronomical Observatories, Chinese Academy of Sciences. R.P.E. is supported by the Chinese Academy of Sciences President’s International Fellowship Initiative, grant No. 2021FSM0004.

\end{acknowledgements}

\label{lastpage}
\bibliography{bibtex}{}

@ARTICLE{2024ChJPh..90..121C,
       author = {{Cao}, Jie and {Xu}, Tingting and {Deng}, Linhua and {Zhou}, Xueliang and {Li}, Shangxi and {Liu}, Yuxia and {Zhou}, Weihong},
        title = "{Pulsar candidate identification using advanced transformer-based models}",
      journal = {Chinese Journal of Physics},
         year = 2024,
        month = aug,
       volume = {90},
        pages = {121-133},
          doi = {10.1016/j.cjph.2024.05.020},
       adsurl = {https://ui.adsabs.harvard.edu/abs/2024ChJPh..90..121C}
}

@ARTICLE{2002AJ....124.1788R,
       author = {{Ransom}, Scott M. and {Eikenberry}, Stephen S. and {Middleditch}, John},
        title = "{Fourier Techniques for Very Long Astrophysical Time-Series Analysis}",
      journal = {\aj},
         year = 2002,
        month = sep,
       volume = {124},
       number = {3},
        pages = {1788-1809},
          doi = {10.1086/342285},
archivePrefix = {arXiv},
       eprint = {astro-ph/0204349},
 primaryClass = {astro-ph},
       adsurl = {https://ui.adsabs.harvard.edu/abs/2002AJ....124.1788R}
}

@ARTICLE{2003ApJ...596.1142C,
       author = {{Cordes}, J.~M. and {McLaughlin}, M.~A.},
        title = "{Searches for Fast Radio Transients}",
      journal = {\apj},
         year = 2003,
        month = oct,
       volume = {596},
       number = {2},
        pages = {1142-1154},
          doi = {10.1086/378231},
archivePrefix = {arXiv},
       eprint = {astro-ph/0304364},
 primaryClass = {astro-ph},
       adsurl = {https://ui.adsabs.harvard.edu/abs/2003ApJ...596.1142C}
}

@ARTICLE{2003ApJ...596..982M,
       author = {{McLaughlin}, M.~A. and {Cordes}, J.~M.},
        title = "{Searches for Giant Pulses from Extragalactic Pulsars}",
      journal = {\apj},
         year = 2003,
        month = oct,
       volume = {596},
       number = {2},
        pages = {982-996},
          doi = {10.1086/378232},
archivePrefix = {arXiv},
       eprint = {astro-ph/0304365},
 primaryClass = {astro-ph},
       adsurl = {https://ui.adsabs.harvard.edu/abs/2003ApJ...596..982M}
}

@ARTICLE{2020MNRAS.497.4654M,
       author = {{Morello}, V. and {Barr}, E.~D. and {Stappers}, B.~W. and {Keane}, E.~F. and {Lyne}, A.~G.},
        title = "{Optimal periodicity searching: revisiting the fast folding algorithm for large-scale pulsar surveys}",
      journal = {\mnras},
         year = 2020,
        month = oct,
       volume = {497},
       number = {4},
        pages = {4654-4671},
          doi = {10.1093/mnras/staa2291},
archivePrefix = {arXiv},
       eprint = {2004.03701},
 primaryClass = {astro-ph.IM},
       adsurl = {https://ui.adsabs.harvard.edu/abs/2020MNRAS.497.4654M}
}

@ARTICLE{2011IJMPD..20..989N,
       author = {{Nan}, Rendong and {Li}, Di and {Jin}, Chengjin and {Wang}, Qiming and {Zhu}, Lichun and {Zhu}, Wenbai and {Zhang}, Haiyan and {Yue}, Youling and {Qian}, Lei},
        title = "{The Five-Hundred Aperture Spherical Radio Telescope (fast) Project}",
      journal = {International Journal of Modern Physics D},
         year = 2011,
        month = jan,
       volume = {20},
       number = {6},
        pages = {989-1024},
          doi = {10.1142/S0218271811019335},
archivePrefix = {arXiv},
       eprint = {1105.3794},
 primaryClass = {astro-ph.IM},
       adsurl = {https://ui.adsabs.harvard.edu/abs/2011IJMPD..20..989N}
}

@ARTICLE{2025ApJS..279...51L,
       author = {{Lian}, Yujie and {Pan}, Zhichen and {Zhang}, Haiyan and {Cao}, Shuo and {Freire}, P.~C.~C. and {Qian}, Lei and {Eatough}, Ralph P. and {Shao}, Lijing and {Ransom}, Scott M. and {Lorimer}, Duncan R. and {Yin}, Dejiang and {Dai}, Yinfeng and {Liu}, Kuo and {Wang}, Lin and {Wang}, Yujie and {Zhang}, Zhongli and {Feng}, Zhonghua and {Li}, Baoda and {Li}, Minghui and {Liu}, Tong and {Li}, Yaowei and {Peng}, Bo and {Pan}, Yu and {Wu}, Yuxiao and {Zhang}, Liyun and {Zhang}, Xingnan and {Jiang}, Peng},
        title = "{The FAST Globular Cluster Pulsar Survey (GC FANS)}",
      journal = {\apjs},
         year = 2025,
        month = aug,
       volume = {279},
       number = {2},
          eid = {51},
        pages = {51},
          doi = {10.3847/1538-4365/ade4ba},
archivePrefix = {arXiv},
       eprint = {2506.07970},
 primaryClass = {astro-ph.HE},
       adsurl = {https://ui.adsabs.harvard.edu/abs/2025ApJS..279...51L}
}

@ARTICLE{2014MNRAS.443.1651M,
       author = {{Morello}, V. and {Barr}, E.~D. and {Bailes}, M. and {Flynn}, C.~M. and {Keane}, E.~F. and {van Straten}, W.},
        title = "{SPINN: a straightforward machine learning solution to the pulsar candidate selection problem}",
      journal = {\mnras},
         year = 2014,
        month = sep,
       volume = {443},
       number = {2},
        pages = {1651-1662},
          doi = {10.1093/mnras/stu1188},
archivePrefix = {arXiv},
       eprint = {1406.3627},
 primaryClass = {astro-ph.IM},
       adsurl = {https://ui.adsabs.harvard.edu/abs/2014MNRAS.443.1651M}
}

@PHDTHESIS{2001PhDT.......123R,
       author = {{Ransom}, Scott Mitchell},
        title = "{New search techniques for binary pulsars}",
       school = {Harvard University, Massachusetts},
         year = 2001,
        month = jan,
       adsurl = {https://ui.adsabs.harvard.edu/abs/2001PhDT.......123R}
}

@ARTICLE{2025ITIF...20.6635L,
       author = {{Liu}, Xiao and {Li}, Mingyuan and {Yu}, Guangsheng and {Wang}, Xu and {Ni}, Wei and {Li}, Lixiang and {Peng}, Haipeng and {Ping Liu}, Ren},
        title = "{BlockFUL: Enabling Unlearning in Blockchained Federated Learning}",
      journal = {IEEE Transactions on Information Forensics},
         year = 2025,
        month = jan,
       volume = {20},
        pages = {6635-6650},
          doi = {10.1109/TIFS.2025.3583109},
archivePrefix = {arXiv},
       eprint = {2402.16294},
 primaryClass = {cs.CR},
       adsurl = {https://ui.adsabs.harvard.edu/abs/2025ITIF...20.6635L}
}

@ARTICLE{2021ApJ...915L..28P,
       author = {{Pan}, Zhichen and {Qian}, Lei and {Ma}, Xiaoyun and {Liu}, Kuo and {Wang}, Lin and {Luo}, Jintao and {Yan}, Zhen and {Ransom}, Scott and {Lorimer}, Duncan and {Li}, Di and {Jiang}, Peng},
        title = "{FAST Globular Cluster Pulsar Survey: Twenty-four Pulsars Discovered in 15 Globular Clusters}",
      journal = {\apjl},
         year = 2021,
        month = jul,
       volume = {915},
       number = {2},
          eid = {L28},
        pages = {L28},
          doi = {10.3847/2041-8213/ac0bbd},
archivePrefix = {arXiv},
       eprint = {2106.08559},
 primaryClass = {astro-ph.HE},
       adsurl = {https://ui.adsabs.harvard.edu/abs/2021ApJ...915L..28P}
}

@ARTICLE{2021RAA....21..143P,
       author = {{Pan}, Zhichen and {Ma}, Xiao-Yun and {Qian}, Lei and {Wang}, Lin and {Yan}, Zhen and {Luo}, Jin-Tao and {Ransom}, Scott M. and {Lorimer}, Duncan R. and {Jiang}, Peng},
        title = "{Three pulsars discovered by FAST in the globular cluster NGC 6517 with a pulsar candidate sifting code based on dispersion measure to signal-to-noise ratio plots}",
      journal = {Research in Astronomy and Astrophysics},
         year = 2021,
        month = aug,
       volume = {21},
       number = {6},
          eid = {143},
        pages = {143},
          doi = {10.1088/1674-4527/21/6/143},
archivePrefix = {arXiv},
       eprint = {2103.14927},
 primaryClass = {astro-ph.HE},
       adsurl = {https://ui.adsabs.harvard.edu/abs/2021RAA....21..143P}
}

@ARTICLE{2024ApJ...969L...7Y,
       author = {{Yin}, Dejiang and {Zhang}, Li-yun and {Qian}, Lei and {Eatough}, Ralph P. and {Li}, Baoda and {Lorimer}, Duncan R. and {Dai}, Yinfeng and {Li}, Yaowei and {Zhang}, Xingnan and {Li}, Minghui and {Su}, Tianhao and {Wu}, Yuxiao and {Pan}, Yu and {Lian}, Yujie and {Liu}, Tong and {Yan}, Zhen and {Pan}, Zhichen},
        title = "{FAST Discovery of Eight Isolated Millisecond Pulsars in NGC 6517}",
      journal = {\apjl},
         year = 2024,
        month = jul,
       volume = {969},
       number = {1},
          eid = {L7},
        pages = {L7},
          doi = {10.3847/2041-8213/ad534e},
archivePrefix = {arXiv},
       eprint = {2405.18228},
 primaryClass = {astro-ph.HE},
       adsurl = {https://ui.adsabs.harvard.edu/abs/2024ApJ...969L...7Y}
}

@ARTICLE{2024ApJ...972...43L,
       author = {{Li}, Baoda and {Zhang}, Li-yun and {Yao}, Jumei and {Yin}, Dejiang and {Eatough}, Ralph P. and {Li}, Minghui and {Li}, Yifeng and {Lian}, Yujie and {Pan}, Yu and {Dai}, Yinfeng and {Li}, Yaowei and {Zhang}, Xingnan and {Su}, Tianhao and {Wu}, Yuxiao and {Liu}, Tong and {Liu}, Kuo and {Wang}, Lin and {Qian}, Lei and {Pan}, Zhichen},
        title = "{Timing and Scintillation Studies of Pulsars in Globular Cluster M3 (NGC 5272) with FAST}",
      journal = {\apj},
         year = 2024,
        month = sep,
       volume = {972},
       number = {1},
          eid = {43},
        pages = {43},
          doi = {10.3847/1538-4357/ad5a82},
archivePrefix = {arXiv},
       eprint = {2406.18169},
 primaryClass = {astro-ph.HE},
       adsurl = {https://ui.adsabs.harvard.edu/abs/2024ApJ...972...43L}
}

@ARTICLE{2024ApJ...974L..23W,
       author = {{Wu}, Yuxiao and {Pan}, Zhichen and {Qian}, Lei and {Ransom}, Scott M. and {Eatough}, Ralph P. and {Wang}, BoJun and {Freire}, Paulo C.~C. and {Liu}, Kuo and {Yan}, Zhen and {Luo}, Jintao and {Zhang}, Liyun and {Li}, Minghui and {Yin}, Dejiang and {Li}, Baoda and {Li}, Yifeng and {Dai}, Yinfeng and {Li}, Yaowei and {Zhang}, Xinnan and {Liu}, Tong and {Pan}, Yu},
        title = "{The Discovery of Three Pulsars in the Globular Cluster M15 with FAST}",
      journal = {\apjl},
         year = 2024,
        month = oct,
       volume = {974},
       number = {2},
          eid = {L23},
        pages = {L23},
          doi = {10.3847/2041-8213/ad7b9e},
archivePrefix = {arXiv},
       eprint = {2312.06067},
 primaryClass = {astro-ph.HE},
       adsurl = {https://ui.adsabs.harvard.edu/abs/2024ApJ...974L..23W}
}

@ARTICLE{2025RAA....25g1001D,
       author = {{Dai}, Yinfeng and {Pan}, Zhichen and {Qian}, Lei and {Zhang}, Liyun and {Yin}, Dejiang and {Li}, Baoda and {Li}, Yaowei and {Wu}, Yuxiao and {Lian}, Yujie},
        title = "{The FAST Discovery of a Millisecond Pulsar M15O (PSR J2129+1210O) Hidden in the Harmonics of M15A (PSR J2129+1210A)}",
      journal = {Research in Astronomy and Astrophysics},
         year = 2025,
        month = jul,
       volume = {25},
       number = {7},
          eid = {071001},
        pages = {071001},
          doi = {10.1088/1674-4527/addc53},
archivePrefix = {arXiv},
       eprint = {2504.16872},
 primaryClass = {astro-ph.HE},
       adsurl = {https://ui.adsabs.harvard.edu/abs/2025RAA....25g1001D}
}

@INPROCEEDINGS{2011ASPC..447..285T,
       author = {{Tauris}, T.~M.},
        title = "{Five and a Half Roads to Form a Millisecond Pulsar}",
    booktitle = {Evolution of Compact Binaries},
         year = 2011,
       editor = {{Schmidtobreick}, L. and {Schreiber}, M.~R. and {Tappert}, C.},
       series = {Astronomical Society of the Pacific Conference Series},
       volume = {447},
        month = sep,
        pages = {285},
          doi = {10.48550/arXiv.1106.0897},
archivePrefix = {arXiv},
       eprint = {1106.0897},
 primaryClass = {astro-ph.HE},
       adsurl = {https://ui.adsabs.harvard.edu/abs/2011ASPC..447..285T}
}
\end{multicols}
\end{document}